\documentclass[12pt,dvips]{article}

\usepackage{rotating}
\usepackage{epsfig}
\usepackage{color}

\newcommand{\T}{\tilde\tau}

\begin{document}

\vbox{  \large
\begin{flushright}
IFT- 02/43 \\
December, 2002 \\
hep--ph/0212388\\
\end{flushright}
\vskip1.5cm
\begin{center}
{\LARGE\bf
Supersymmetry Parameter Analysis}\\[1cm]

{\Large Jan Kalinowski}\\[5pt]{\it 
Instytut Fizyki Teoretycznej, Uniwersytet Warszawski\\[2mm] 
           PL--00681 Warsaw,     Poland}\\

\vskip2cm
\thispagestyle{empty}

{\bf Abstract}\\[1pc]

\begin{minipage}{15cm}
  Supersymmetric particles can be produced copiously at future
  colliders. From the high-precision data taken at $e^+e^-$ linear
  colliders, TESLA in particular, and combined with results from LHC,
  and CLIC later, the low-energy parameters of the supersymmetric
  model can be determined.  Evolving the parameters from the
  low-energy scale to the high-scale by means of renormalization group
  techniques the  fundamental supersymmetry parameters at the high
  scale, GUT or Planck, can be reconstructed to reveal the origin of
  supersymmetry breaking.
\end{minipage}  \\
\vskip2.5cm
Invited talk at \\
The 10th International Conference on Supersymmetry and \\ 
Unification of Fundamental Interactions (SUSY02) \\
17--23 June 2002, DESY Hamburg, Germany\\
\end{center}
}
\vfill
\clearpage

\setcounter{page}{1}

\begin{center}
{\large{\bf SUPERSYMMETRY PARAMETER ANALYSIS \\[3mm]
            Jan Kalinowski\\[5mm] }}
 Instytut Fizyki Teoretycznej, Uniwersytet Warszawski\\[2mm] 
           PL--00681 Warsaw,     Poland
\end{center}
\bigskip
\bigskip
\bigskip
\vskip 2cm

\begin{abstract}
  Supersymmetric particles can be produced copiously at future
  colliders. From the high-precision data taken at $e^+e^-$ linear
  colliders, TESLA in particular, and combined with results from LHC,
  and CLIC later, the low-energy parameters of the supersymmetric
  model can be determined.  Evolving the parameters from the
  low-energy scale to the high-scale by means of renormalization group
  techniques the  fundamental supersymmetry parameters at the high
  scale, GUT or Planck, can be reconstructed to reveal the origin of
  supersymmetry breaking.
\end{abstract}
%

\section{Introduction}

\def\beq{\begin{equation}} 
\def\eeq{\end{equation}} 
\def\bea{\begin{eqnarray}}  
\def\eea{\end{eqnarray}}  
\def\bq{\begin{quote}}  
\def\eq{\end{quote}}  
  
\def\beqa{\begin{eqnarray}}  
\def\eeqa{\end{eqnarray}}  
\def\be{\begin{equation}}  
\def\ee{\end{equation}}  
\def\beq{\begin{equation}}  
\def\eeq{\end{equation}}  
  
\def\noin{\noindent}  
\def\grad{{ \nabla}}  
\def\pa{\partial}  
\def\kaps{{\kappa}^{2}}  
\def\Melev{ M^{11}}  

\def\cp{{\cal P}}  
\def\cl{{\cal L}} 
\def\cf{{\cal F}} 
\def\tsi{\tilde{\sigma}}  
\def\tga{\tilde{\gamma}}  
\def\r2{\sqrt{2}}  
\def\ra{\rightarrow}
\def\bi{\begin{itemize}}  
\def\ei{\end{itemize}}  
\def\i{\item}  
  
\def\ov{\overline}  
\def\nn{\nonumber \\}  
  
\def\mh{\hat{\mu}}  
\def\nh{\hat{\nu}}  
\def\lh{\hat{\lambda}}  
\def\fip{\phi_5}  
\def\lc{{\cal L}}  
\def\ca{{\cal A}}  
\def\var{\vartheta}  
\newcommand{\imag}{\Im {\rm m}}
\newcommand{\real}{\Re {\rm e}}
\newcommand{\smu}{\tilde{\mu}}
\newcommand{\se}{\tilde{e}}
\newcommand{\eR}{e_{\rm R}}
\newcommand{\eL}{e_{\rm L}}
\newcommand{\seR}{\tilde{e}_{\rm R}}
\newcommand{\seL}{\tilde{e}_{\rm L}}
\newcommand{\smR}{\tilde{\mu}_{\rm R}}
\newcommand{\smL}{\tilde{\mu}_{\rm L}}
\newcommand{\slpt}{\tilde{l}}
\newcommand{\schi}{\tilde{\chi}}
\newcommand{\mn}[1]{m_{\tilde{\chi}^0_{#1}}}
\newcommand{\mnsq}[1]{m^2_{\tilde{\chi}^0_{#1}}}

Despite the lack of direct experimental evidence\footnote{The current
experimental status of low-energy supersummetry is discussed in
\cite{status}.} 
for supersymmetry (SUSY), the concept of symmetry between bosons and
fermions has so many attractive features that the supersymmetric
extension of the Standard Model is widely considered as a most natural
scenario.  It protects the electroweak scale from destabilizing
divergences, leads to the unification of gauge couplings,
accommodates a large top quark mass, provides a natural candidate for
dark matter, and decouples from precision measurements.  Exact
supersymmetry is a fully predictive framework: to each known particle
it predicts the existence of its superpartner which differs in the
spin quantum number by 1/2, and fixes their couplings without
introducing new parameters.  Thus the Minimal Supersymmetric extension
of the Standard Model (MSSM) encompasses spin 1/2 partners of the gauge
and Higgs bosons, called gauginos and higgsinos, and spin 0 companions
of leptons and quarks, called sleptons and squarks.

If realized in nature, supersymmetry (SUSY) must be a broken
symmetry since until now  
none of the superpartners have been found. The construction of a
viable mechanism of SUSY breaking, however, is a difficult issue. It is
impossible to construct a realistic breaking scenario with the
particle content of the MSSM \cite{massr}. 
Therefore the origin of SUSY breaking is
usually assumed to take place in a ``hidden sector'' of particles
which have no 
direct couplings to the MSSM particles and the supersymmetry breaking
is ``mediated'' from the hidden to the visible sector. This opens up a
variety of possible scenarios of SUSY breaking and its mediation, and  
in fact many models have been proposed: gravity-mediated,
gauge-mediated, anomaly-mediated,
gaugino-mediated etc. Each model is characterized by a few
parameters (usually defined at a high scale) and leads to different
phenomenological consequences.   

With all the different  SUSY models proposed in the past, the best 
is to keep an open mind and  parameterize the
breaking of SUSY by the most general explicit breaking terms in the
Lagrangian. The structure of the breaking terms is constrained by the
gauge symmetry and the requirement of stabilization against radiative
corrections from higher scales. This leads to a set of soft-breaking
terms \cite{softbr}, which include \\ 
(i) gaugino mass terms for bino
$\tilde{B}$, wino $\tilde{W}^j$ $[j=$1--3] and gluino $\tilde{g}^i$
$[i=$1--8] 
\beq {\textstyle \frac{1}{2}} M_1
\,\overline{\tilde{B}}\,\tilde{B} \ + \ {\textstyle \frac{1}{2}} M_2
\,\overline{\tilde{W}}^i \,\tilde{W}^i \ + \ {\textstyle \frac{1}{2}}
M_3 \,\overline{\tilde{g}}^i \,\tilde{g}^i \ , 
\eeq 
(ii) trilinear ($A^i$) and bilinear ($B$) scalar 
couplings (generation indices are understood) 
\beq A_u H_2 \tilde{Q}
\tilde{u}^c + A_d H_1 \tilde{Q} \tilde{d}^c + A_l H_1 \tilde{L}
\tilde{l}^c - \mu B H_1 H_2 
\eeq 
(iii) and squark and slepton mass
terms \beq m_{\tilde{Q}}^2 [\tilde{u}^*_L\tilde{u}_L
+\tilde{d}^*_L\tilde{d}_L] + m_{\tilde{u}}^2 \tilde{u}^*_R \tilde{u}_R
+ m_{\tilde{d}}^2 \tilde{d}^*_R \tilde{d}_R+
 \ \cdots
\eeq
where the ellipses stand for the soft mass terms for sleptons 
and Higgs bosons (tilde denotes the superpartner). 
The above parameters can be complex with nontrivial
CP violating phases \cite{Nath}. 
The stability of quantum corrections
implies that at least some of the superpartners should be relatively light,
with masses around 1 TeV, and thus within the reach of present or next
generation of high-energy colliders.    

As a consequence of the most general soft-breaking terms, a large
number of parameters is introduced. The unconstrained low-energy MSSM
has some 100 parameters resulting in a rich spectroscopy of states and
complex phenomenology of their interactions.   One should
realize, that the low-energy parameters are of two distinct
categories. The first one includes all the gauge and Yukawa couplings
and the higgsino mass parameter $\mu$. They are related by exact
supersymmetry which is crucial for the cancellation of quadratic
divergences.  For example, at tree-level the $qqZ$,
$\tilde{q}\tilde{q}Z$ gauge and $q\tilde{q}\tilde{Z}$ Yukawa couplings
have to be equal.  
The relations among these parameters 
(with calculable radiative corrections) have to be
confirmed experimentally; if not -- the supersymmetry is excluded.
The second category encompasses all soft-supersymmetry breaking
parameters: higgsino, gaugino and sfermion masses and mixing, and
the trilinear couplings. 

If supersymmetry is
detected at a future collider it will be a matter of days to discover
all kinetically accessible supersymmetric particles. However, it will 
be an enormous task to
investigate the masses, couplings and quantum numbers of the
superpartners and many measurements and considerable ingenuity will be
needed to reconstruct a complete low-energy theory.
The experimental program to search for and explore SUSY at present and
future colliders has to include the following points:\\[1mm]
\hspace*{10mm} (a) discover supersymmetric particles and measure their quantum
   numbers to prove\\
\hspace*{16mm} that they are {\it the} superpartners of standard
   particles,\\
\hspace*{10mm} (b) measure their masses, mixing angles and couplings,\\
\hspace*{10mm} (c)
   determine the low-energy Lagrangian parameters,\\
\hspace*{10mm} (d) verify the relations among them in order to distinguish
   between various SUSY\\
\hspace*{16mm} breaking models.

\begin{figure}[th]
\begin{minipage}[b]{11cm}
 \epsfig{file=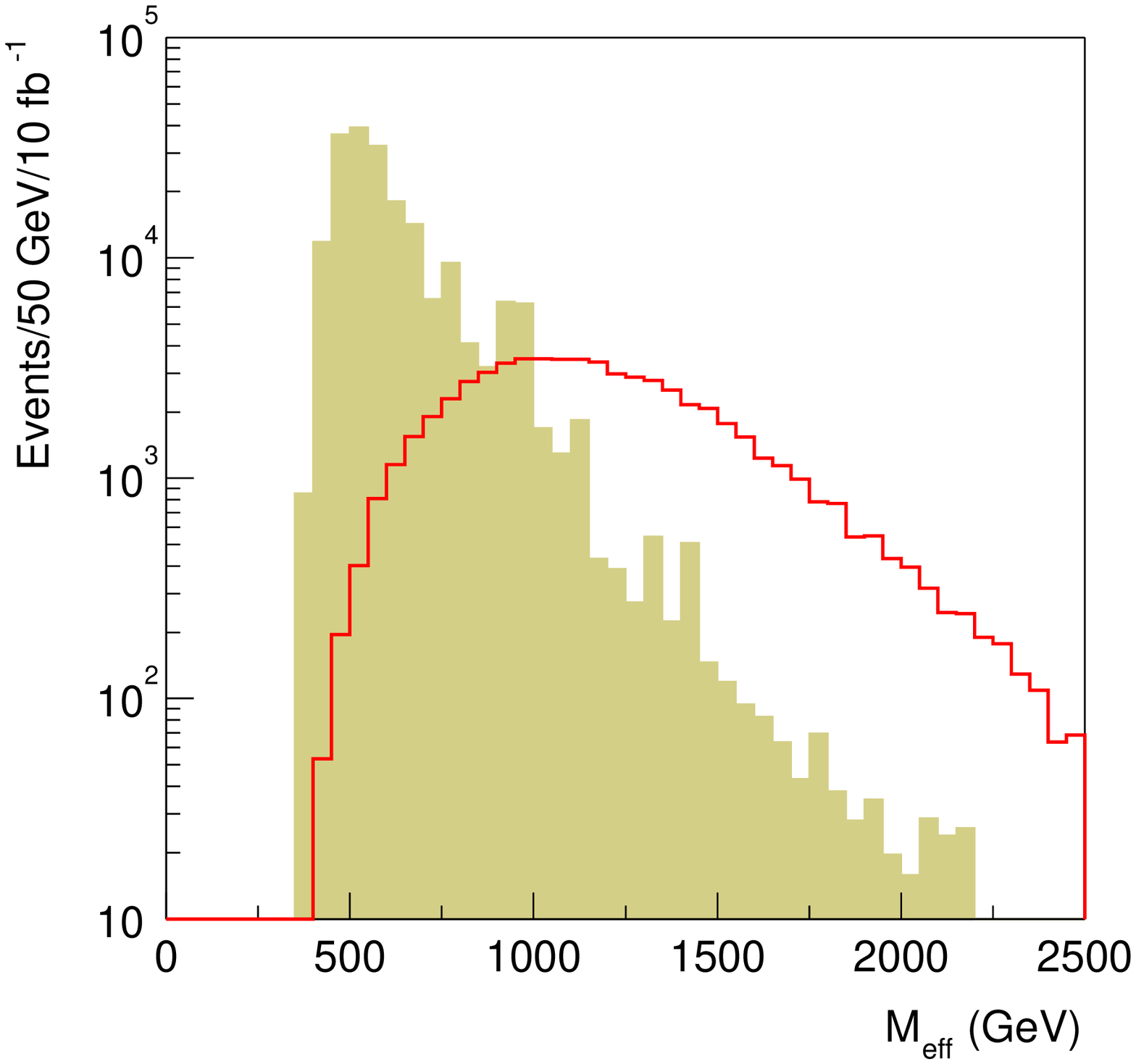,height=5.5cm,width=5.5cm}
\epsfig{file=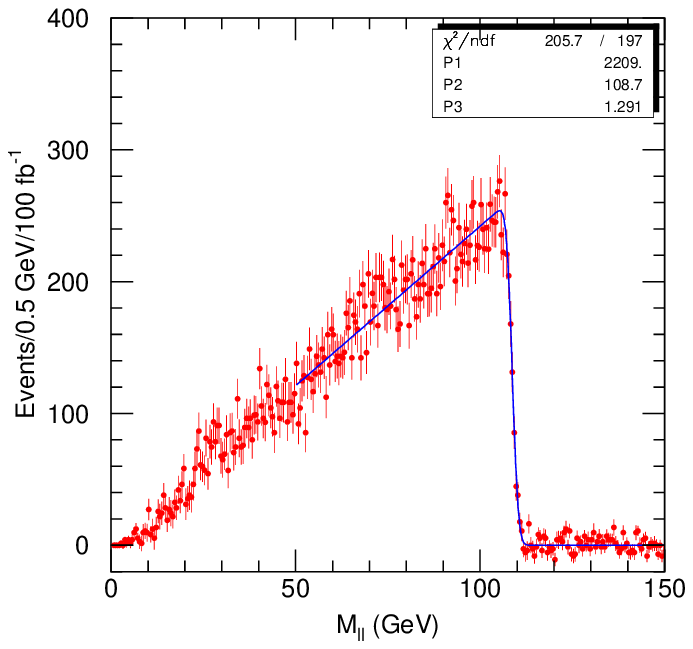,height=5cm,width=5cm}
\end{minipage}%
\begin{minipage}[b]{5cm}
\caption{Left: Typical $M_{eff}$ spectrum. Right: The dilepton
invariant mass distribution. From \cite{SusyLHC}.}
\end{minipage}%
\label{fig:lhc}
\end{figure}

It is important to realize that all low-energy SUSY parameters should
be measured independently of any theoretical assumptions.  In this
respect the concept of a high-energy $e^+e^-$ linear collider
\cite{LC} is of particular interest since it opens up a possibility of
precision measurements of supersymmetric particle properties.  An
intense activity during last few years on $e^+e^-$ collider physics
has convincingly demonstrated the advantages and benefits of such a
machine and its complementarity to the Large Hadron Collider (LHC).
Many studies have shown that the LHC can cover a mass range for SUSY
particles up to $\sim$ 2 TeV, in particular for squarks and gluinos
\cite{SusyLHC}.  Early indication of SUSY can be provided by an excess
in $M_{eff}=E^T_{miss} + p^T_1+p^T_2 +\ldots$, an example is shown in
Fig.\ref{fig:lhc}, 
where $E^T_{miss}$ is due to the stable lightest SUSY particles
(LSP) escaping detection, and $p^T_i$ are transverse momenta of jets.
Sparticles from squark and gluino decays can be accessed if the
SUSY decays are distinctive. The problem however is that many
different sparticles will be accessed at once with the heavier ones
cascading into the lighter which will in turn cascade further leading
to a complicated picture.  Simulations for the extraction of
parameters have been attempted for the LHC and demonstrated that some
of them can be extracted with a good precision.  Identifying
particular decay channels and measuring the endpoints, for example in the
dilepton invariant mass as shown in Fig.\ref{fig:lhc}, 
the mass differences of SUSY
particles can be determined very precisely. If enough channels are
identified and measured, the masses can be determined without any
model assumptions. With a large amount of information available from
the production of squarks and gluinos and their main decay products,
theoretical interpretation can be possible in favorable models.

But if all proposed theoretical models of the SUSY breaking turn out
to be  wrong, the concurrent running of an $e^+e^-$ LC will be very
much welcome. It will provide information complementary to that from
the LHC.   Thanks to\\[1mm]
\hspace*{10mm} $\diamond$ clean final state environment,\\ 
\hspace*{10mm} $\diamond$ tunable energy,\\
\hspace*{10mm} $\diamond$ polarized incoming beams,\\ 
\hspace*{10mm} 
$\diamond$ and a possibility of additional modes: $e^-e^-$, $e\gamma$ and
$\gamma\gamma$\\[1mm] 
precise determination of masses, couplings, quantum numbers, mixing
angles and CP phases will be possible at $e^+e^-$ colliders.  As I
will illustrate with some examples below, 
this will allow us a model independent
reconstruction of the low-energy SUSY parameters to be performed and,
hopefully, connect the low-scale phenomenology with the high-scale
physics \cite{Kane}.

\section{Reconstruction of low-energy SUSY parameters}
In contrast to many earlier analyzes, we will
not elaborate on global fits 
but rather we will discuss attempts at
``measuring'' the fundamental Lagrangian parameters.  
Generically such attempts are performed in two steps \cite{jkkm}\\[1mm]
\hspace*{10mm} $A$: from the observed quantities: cross sections, asymmetries 
    etc. determine the \\
\hspace*{16mm} physical parameters:
    the masses, mixing and couplings of sparticles, \\
\hspace*{10mm} $B$: from the physical parameters  extract
    the Lagrangian parameters: $M_i$, $\mu$, $\tan\beta$, \\
\hspace*{16mm} $A^u$, $m_{\tilde{Q}}$ etc.

To deal with so many parameters, a clear strategy is needed. An
attractive possibility would be to \\[1mm]
\hspace*{10mm} $\diamond$ start with charginos which depend only on   $M_2$,
$\mu$, $\tan\beta$, \\
\hspace*{10mm} $\diamond$ add neutralinos which depend in addition on
$M_1$,\\  
\hspace*{10mm} $\diamond$ include sleptons which bring in 
$m_{\tilde{l}}$, $A_l$,\\ 
\hspace*{10mm} $\diamond$ and finally squarks and gluinos to determine $ 
m_{\tilde{q}}$,
$A_q$ and $M_3$  \\[1mm]
to reconstruct at tree level the basic structure
of SUSY Lagrangian. 
In reality 
it might be  difficult to separate a specific sector  
({\it e.g.}  sleptons enter
via t-channel in  the chargino production processes),  
many production channels can simultaneously be open, and 
SUSY constitutes an important background to SUSY
processes. In addition  sizable  loop corrections  will  mix all
sectors.  Precision measurements will require 
loop corrections to the masses and mixing angles, the 
finite decay width effects and  
loop corrections to the production and decay processes
to be included  for final  global analyzes of all data.
Some one-loop results are already available: 
for the current status we refer to \cite{Majerotto}. 
\subsection{The chargino sector}

\begin{figure}[th!]
\begin{minipage}[b]{7.5cm}
\epsfig{file=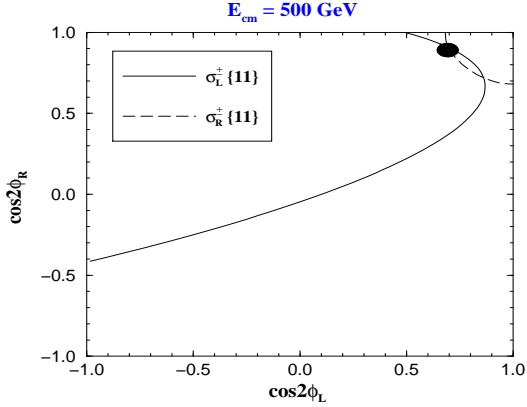, width=7cm, height=5.5cm}
\end{minipage}%
\begin{minipage}[b]{6.5cm}
\caption{Contours of the light chargino production cross sections with
polarized electron beams in the plane 
$[\cos2\phi_L,\cos2\phi_R]$. From \cite{neut}.}
\label{fig:char}
\end{minipage}
\end{figure}

The mass matrix of the wino and charged higgsino, 
after the gauge symmetry breaking,
is nondiagonal 
\begin{eqnarray} 
{\cal M}_C=\left(\begin{array}{cc}
   M_2                &      \sqrt{2}m_W\cos\beta  \\
   \sqrt{2}m_W\sin\beta  &             |\mu|{\rm e}^{i\Phi_\mu}   
                  \end{array}\right)\ 
\end{eqnarray}
It can be diagonalized by  two unitary matrices acting on left- and
right-chiral states 
\begin{eqnarray} 
 U_L=\left(\begin{array}{cc}
  c_L & s_L^* \\ -s_L & c_L
 \end{array}\right), \qquad
 U_R=\left(\begin{array}{cc}
 {\rm e}^{i\gamma_1} & 0 \\ 0 & {\rm e}^{i\gamma_2}
 \end{array}\right)
 \left(\begin{array}{cc}
 c_R  & s_R^* \\  -s_R  & c_R  \end{array}\right)  \label{chardiag}
\end{eqnarray}
with $ c_{L,R}=\cos\phi_{L,R}$, $ s_{L,R}={\rm e}^{i\beta_{L,R}}
\sin\phi_{L,R}  $,  
which involve  two mixing angles $\phi_{L,R}$ and three CP 
phases $\beta_{L,R}$ and 
$\gamma_1-\gamma_2$  
The mass eigenstates, called charginos, are mixtures of wino and higgsino
with the masses and mixing angles given  by
\begin{eqnarray}
&&m^2_{\tilde{\chi}^\pm_{1,2}}
   =\frac{1}{2}\left[M^2_2+|\mu|^2+2m^2_W\mp { \Delta_C}\right]
\label{massch}\\
&&\cos 2\phi_{L,R}=-\left[M_2^2-|\mu|^2\mp 2m^2_W\cos 2\beta\right]/
{ \Delta_C}\label{mixch}
\end{eqnarray}
where $\Delta_C=[(M^2_2-|\mu|^2)^2+4m^4_W\cos^2 2\beta
              +4m^2_W(M^2_2+|\mu|^2)
+8m^2_WM_2|\mu|
               \sin2\beta\cos{\Phi_\mu}]^{1/2}$.
Experimentally the chargino 
masses can be measured very precisely either by  threshold scans or 
in continuum above the threshold \cite{Martyn}. Since the chargino
production cross sections are simple binomials of $\cos2\phi_{L,R}$,
              see Fig.\ref{fig:char}, the 
mixing angles can be determined in a model independent way 
using  polarized electron beams   \cite{char, neut}. 

Based on this high-precision information, the fundamental SUSY
parameters can be extracted in analytic form. 
Inverting eqs.(\ref{massch},\ref{mixch}) one finds  to lowest
order:
\begin{eqnarray}
M_2&=&M_W[\Sigma - \Delta[\cos2\phi_R+\cos2\phi_L]]^{1/2}\label{eq:m2}\\
\left|\mu\right|&=&M_W[\Sigma + \Delta[\cos2\phi_R+\cos2\phi_L]]^{1/2}
\label{eq:mu}\\
\cos\Phi_\mu&= &[ \Delta^2
                   -(M^2_2-\mu^2)^2-4m^2_W(M^2_2+\mu^2) \nonumber \\
 & &                   -4m^4_W\cos^2 2\beta]/8 m_W^2M_2|\mu|\sin2\beta 
\label{eq:cos}\\
\tan\beta&=&\left[\frac{1+\Delta (\cos 2\phi_R-\cos 2\phi_L)}
           {1-\Delta (\cos 2\phi_R-\cos 2\phi_L)}\right]^{1/2} 
\label{eq:tan}
\end{eqnarray}
where $\Delta =\Delta_C/4M^2_W=
= (m^2_{\tilde{\chi}^\pm_2}-m^2_{\tilde{\chi}^\pm_1})/4M^2_W$
and 
$\Sigma =  (m^2_{\tilde{\chi}^\pm_2}+m^2_{\tilde{\chi}^\pm_1})/2M^2_W -1$.

If both $m_{\tilde{\chi}^\pm_1}$ and $m_{\tilde{\chi}^\pm_2}$ 
can be measured, the fundamental
parameters (\ref{eq:m2}-\ref{eq:tan})
can be extracted unambiguously. 
However, if $\tilde{\chi}^\pm_2$ happens to be beyond the kinematical
reach at an early stage of the LC, 
it depends on the 
CP properties of the higgsino sector whether they can be determined or 
not in the light chargino system alone \cite{neut}

{\it (i)}\, If the higgsino sector is 
     CP invariant, $\cos\Phi_\mu =\pm 1$
     can be exploited to determine 
     $m_{\tilde{\chi}^\pm_2}$ up to at most 
     a two--fold ambiguity.
     This ambiguity can be resolved if other observables
     can be measured, e.g.  the mixed--pair $\tilde{\chi}^0_1
     \tilde{\chi}^0_2$ production cross sections.

{\it (ii)}\, In a CP non--invariant theory the parameters in
     eqs.(\ref{eq:m2}--\ref{eq:tan}) remain dependent on the unknown
     heavy chargino mass $m_{\tilde{\chi}^\pm_2}$.  Two trajectories
     in the plane $[\cos2\phi_L,\cos2\phi_R]$ are generated (and
     consequently in the $\{M_2, \mu; \tan\beta\}$ space),
     parametrized by $m_{\tilde{\chi}^\pm_2}$ and classified by the
     two possible signs of $\sin\Phi_\mu$. The analysis of the two
     light neutralino states $\tilde{\chi}^0_1$ and $\tilde{\chi}^0_2$
     can be used to predict the heavy chargino mass
     $m_{\tilde{\chi}^\pm_2}$ in the MSSM. Therefore we will now
     discuss


\subsection{The neutralino sector}

The mass matrix  of the 
$(\tilde{B},\tilde{W}^3,\tilde{H}^0_1,\tilde{H}^0_2)$ is symmetric but
nondiagonal
\begin{eqnarray} M_N=
\left(\begin{array}{cccc}
  M_1       &      0          &  -m_Z c_\beta s_W  & m_Z s_\beta s_W
  \\
   0        &     M_2         &   m_Z c_\beta c_W  & -m_Z s_\beta
  c_W\\ 
-m_Z c_\beta s_W & m_Z c_\beta c_W &       0       &     -\mu
  \\ 
 m_Z s_\beta s_W &-m_Z s_\beta c_W &     -\mu      &       0
                  \end{array}\right)\
\end{eqnarray}
where $M_1=|M_1|\,\,{\rm e}^{i\Phi_1}$, $\mu=|\mu|\,\,{\rm e}^{i\Phi_\mu} $.
The mass eigenstates, neutralinos, are obtained by  
the $4\times4$ diagonalization matrix $N$, which  
is parameterized by 6 angles and 10 phases as
\begin{eqnarray}
N&=&
 {\sf diag}\left\{{\rm e}^{i\alpha_1},\, 
                   {\rm e}^{i\alpha_2},\,
                   {\rm e}^{i\alpha_3},\,
                   {\rm e}^{i\alpha_4}\,\right\} 
{\sf R}_{34}\, {\sf R}_{24}\,{\sf R}_{14}\,{\sf R}_{23}\,{\sf
  R}_{13}\, {\sf R}_{12} 
\end{eqnarray}
where ${\sf R}_{jk}$ are  $4\times4$ matrices describing 2-dim complex 
rotations in the \{jk\} plane (of the form analogous to $U_L$ in
eq.(\ref{chardiag}) and defined  in terms of  
 ($ cos\theta_{jk}$, $\sin\theta_{jk}\, {\rm e}^{i\delta_{jk}})$).

The CP is conserved in the neutralino sector if $\delta_{ij}=0$  and
$\alpha_i=0$. The  unitarity constraints can conveniently be
formulated in terms of unitarity quadrangles built
up by \\[1mm]
\hspace*{10mm} $\diamond$  the links $N_{ik}N^*_{jk}$  
connecting two {\it rows} $i$ and $j$\\
\hspace*{10mm} $\diamond$ the links $N_{ki}N^*_{kj}$ connecting two {\it
columns}  $i$ and $j$.  \\[1mm]
Unlike in the CKM or MNS cases of quark and lepton mixing, the 
orientation of all quadrangles is physical \cite{neut}.

\begin{figure}[h!]
\begin{minipage}[b]{10.5cm} 
 \epsfig{file=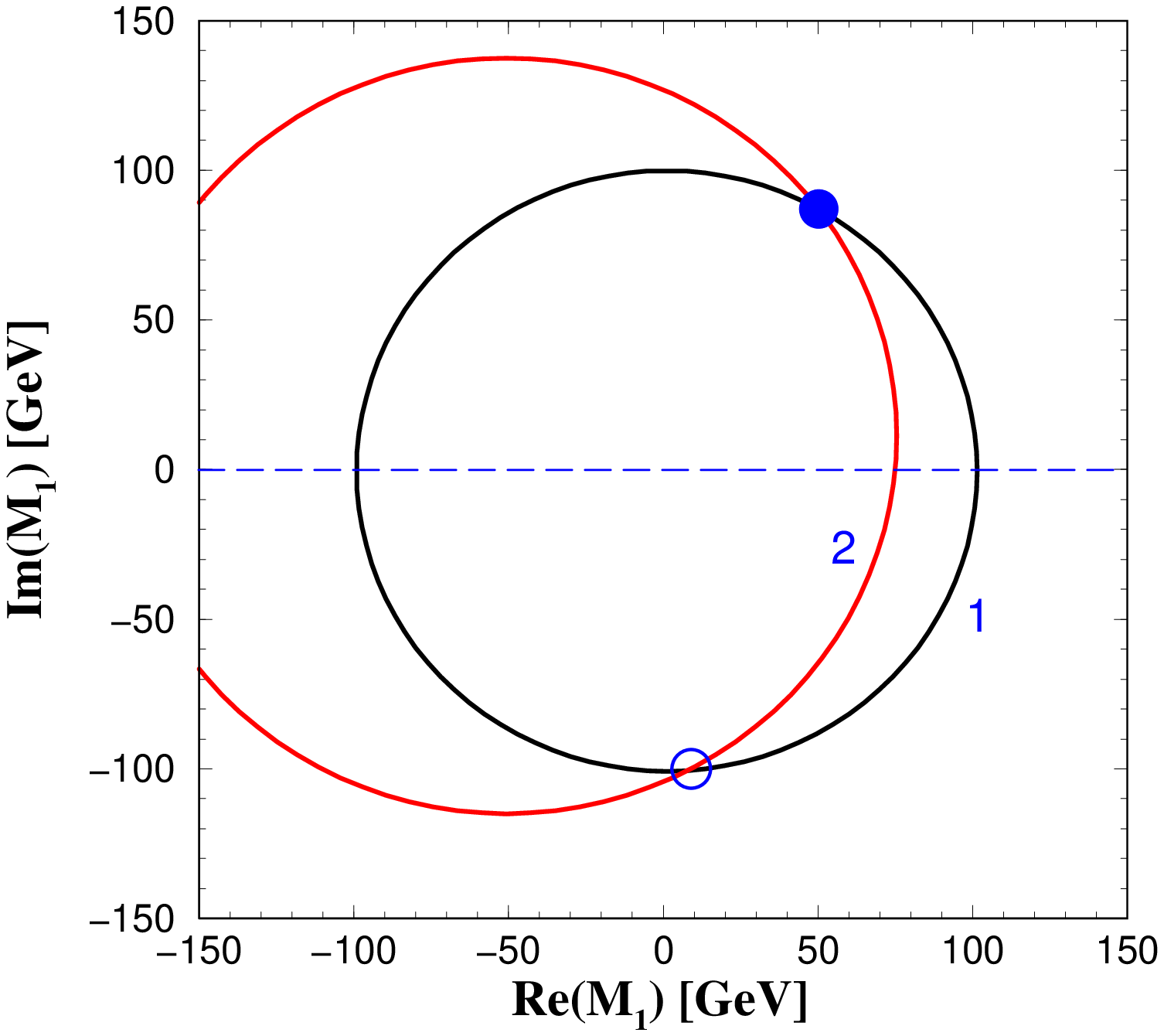,height=4cm,width=4.5cm} \hspace{5mm}
 \epsfig{file=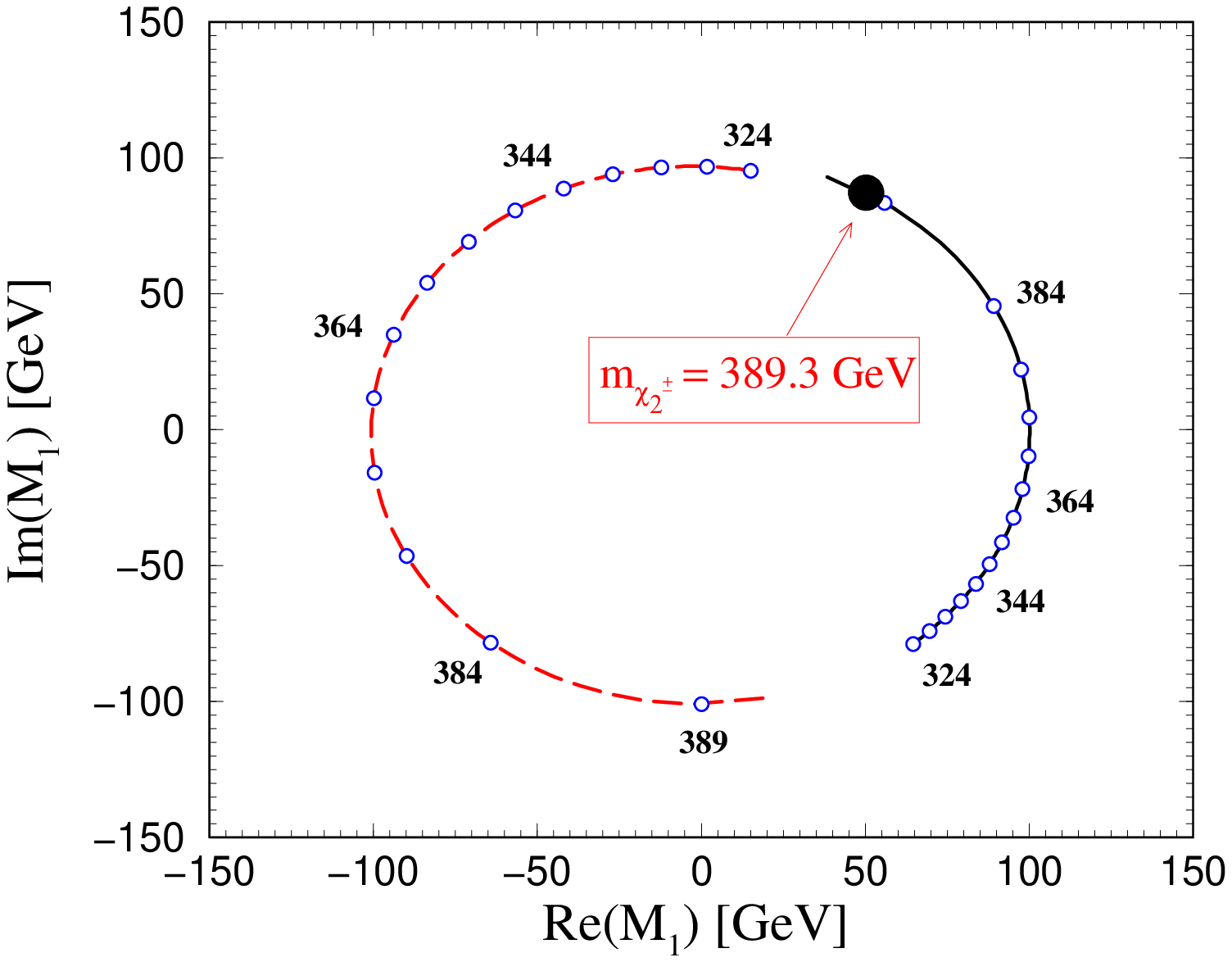,height=4cm,width=4.5cm} 
\end{minipage}%
\begin{minipage}[b]{5cm}
\caption{Left: Contour lines for two light neutralino masses in the 
($\real{M_1}$,  
 $\imag{M_1}$) plane. Right:
The migration with $m_{\tilde{\chi}^\pm_2}$ of two crossing points. 
From \cite{neut}.
\label{fig:circle}
}
\end{minipage}%
\end{figure}

To resolve the light chargino case
in the CP-violating scenario {\it (ii)}, we note that each 
neutralino mass satisfies the characteristic equation
 \begin{eqnarray}
m_{\tilde{\chi}^0_i}^8-a\, m_{\tilde{\chi}^0_i}^6
+b\, m_{\tilde{\chi}^0_i}^4-c\, m_{\tilde{\chi}^0_i}^2+d=0 
\end{eqnarray}
where 
$a$, $b$, $c$ and $d$  are  binomials of
{\ $\real{M_1}$}  
and {\ $\imag{M_1}$}.
Therefore the
equation for each $m^2_{\tilde{\chi}^0_i}$ has the form
\begin{eqnarray}
(\real{M_1})^2+(\imag{M_1})^2+ u_i\, \real{M_1}+ v_i\, \imag{M_1}
 = w_i 
\label{eq:Mphase}
\end{eqnarray}
i.e. each neutralino mass  defines a circle in the $\{\real M_1,
\imag M_1\}$ plane. 
With  two light neutralino masses two crossing points in the ($\real{M_1}$,  
 $\imag{M_1}$) plane are generated, as seen in the left panel of 
Fig.\ref{fig:circle}.

Since from the chargino sector  
$\{M_2,\mu\; \tan\beta\}$ are 
parameterized by unknown $m_{\tilde{\chi}^\pm_2}$, 
the crossing points will migrate with $m_{\tilde{\chi}^\pm_2}$, right
panel of Fig.\ref{fig:circle}. 
One can use the measured cross section  for
$\tilde{\chi}^0_1\tilde{\chi}^0_2$ to select a { unique}
solution for { $M_1$} and predict the 
heavy chargino mass. If the LC runs concurrently with the LHC, 
the LHC experiments may be able to verify the predicted value of   
$m_{\tilde{\chi}^\pm_2}$.

If the machine energy is above the heavy charginos and neutralinos, 
one can\\[1mm] 
\hspace*{10mm} $\diamond$ 
study the threshold behavior of non-diagonal neutralino 
pair production to check\\ \hspace*{14mm} for a 
clear signal of nontrivial CP phases,\\   
\hspace*{10mm} $\diamond$ 
measure the normal neutralino polarization which provides  
a unique probe of\\ \hspace*{14mm} $\alpha_i$ (Majorana) CP phases,\\
\hspace*{10mm} $\diamond$ exploit the sum rules to verify the  closure of the
chargino and neutralino sectors,\\
\hspace*{10mm} $\diamond$ 
analyze the SUSY relations between Yukawa and gauge couplings,\\
\hspace*{10mm} $\diamond$ 
extract information on sleptons exchanged in the t-channels etc.\\[1mm]
For more details, we refer to \cite{conneu}.
\subsection{ The sfermion sector}
 
The  sfermion mass matrix in the $(\tilde{f}_L,\tilde{f}_R)$ basis is
given as 
\begin{eqnarray}
\left(\begin{array}{cc}
    {\ M^2_{\tilde{F_L}}}
+m^2_Z\cos2\beta(I_{f_i}-Q_f 
\sin^2\theta_W) +m_f^2     & m_f  ({\ A_f^*} -{\
\mu}(\cot\beta)^{2I_{f}})  \\
  m_f ({\ A_f} -{\
\mu}^*(\cot\beta)^{2I_{f}})           
&     {\ M^2_{\tilde{f_R}}}
-m^2_Z Q_f \cos2\beta  
\sin^2\theta_W +m_f^2 
\end{array}\right)\ 
\end{eqnarray}
where 
$M^2_{\tilde{F_L}}$, $M^2_{\tilde{f_R}}$ and $A_f$ are slepton soft
SUSY breaking parameters.   
The mass eigenstates are defined 
\begin{eqnarray*} 
 \left(\begin{array}{c} \tilde{f}_1\\ \tilde{f_2} \end{array}\right)
 =\left(\begin{array}{cc}
           {\rm e}^{i\alpha_f}  \cos\theta_f & \sin\theta_f \\
           -\sin\theta_f & {\rm e}^{-i\alpha_f} \cos\theta_f
             \end{array}\right)
\left(\begin{array}{c} \tilde{f}_L\\ \tilde{f_R} \end{array}\right)
\end{eqnarray*}
The 
$(\tilde{f}_L,\tilde{f}_R)$ mixing is important if $|m^2_{\tilde{f}_L}-
m^2_{\tilde{f}_R}|\leq |a_f m_f|$. 
Therefore for the first and second generation sfermions the mixing is usually
neglected. 

\newcommand{\Eslash}{{\not{\!\!E}}}
\begin{figure}[h!]
\begin{minipage}[b]{11cm} 
$e^+e^- \to \seR^+\seR^- \to e^+e^- + \Eslash$ \hspace{6mm}
 $e^-e^- \to \seR^-\seR^- \to e^-e^- + \Eslash$ \\[.5ex]
\epsfig{file=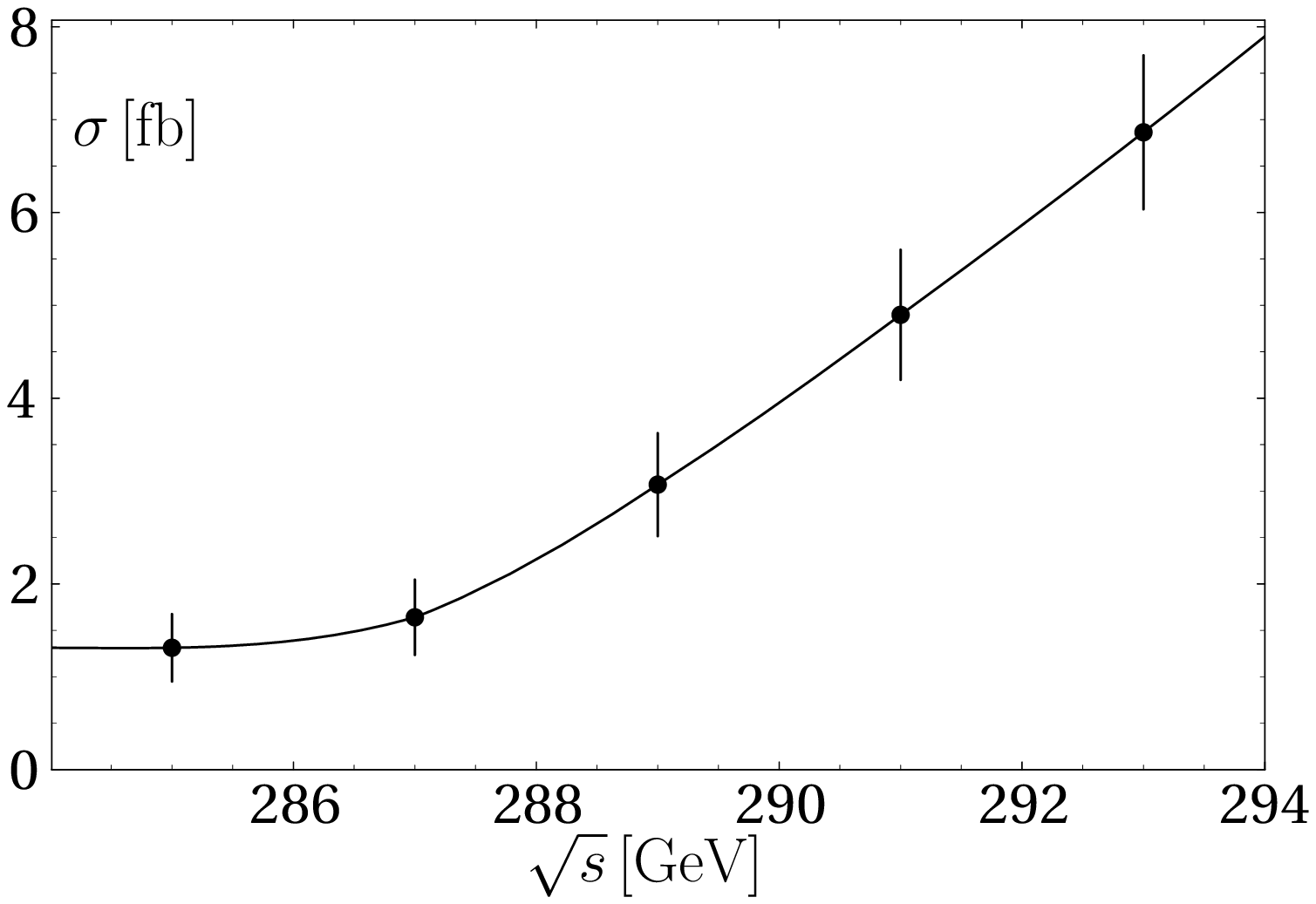,width=5.5cm,
	viewport=30 0 472 290, clip=false}
\epsfig{file=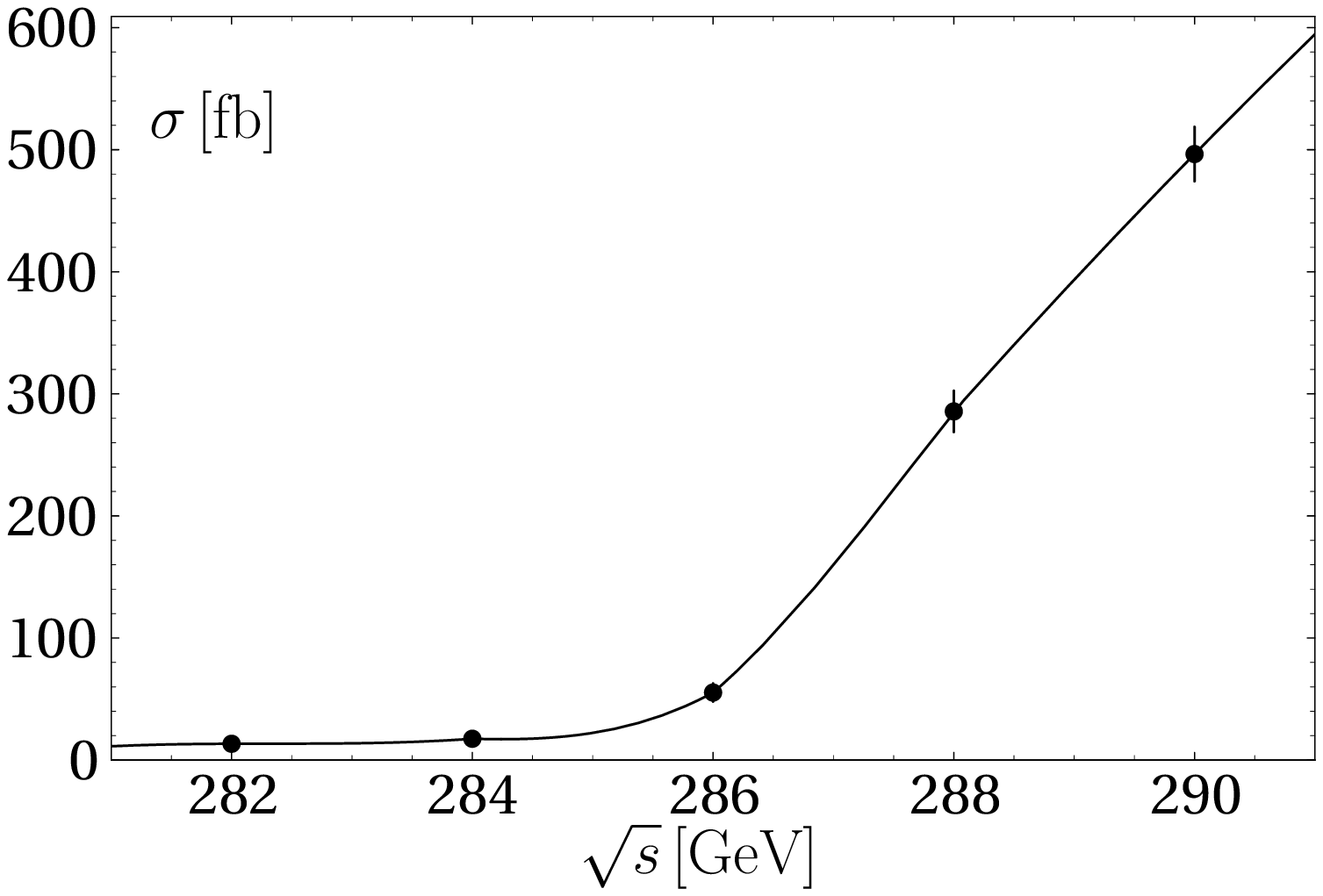,width=5.3cm,
	viewport=43 0 472 290, clip=false}
\end{minipage}%
\hspace{3mm}\begin{minipage}[b]{4.5cm}
\caption{Threshold excitation curves for $\tilde{e}_R$ pair
production. Errors for  $\int L=10 
{\rm \ fb}^{-1}$ in $e^+e^-$ and $1 
{\rm \ fb}^{-1}$ in $e^-e^-$ per scan point. From \cite{Freitas}.}
\label{fig:thrsel}
\end{minipage}
\end{figure}
The slepton masses can be measured at a high luminosity 
$e^+e^-$ collider by scanning 
the pair production near threshold \cite{Martyn}. Since the
expected experimental accuracy is of ${\cal O}(100)$ MeV, it is necessary to
incorporate effects beyond leading order in the theoretical predictions
\cite{Freitas}. The non-zero widths of the sleptons, which considerably affect
the cross-sections near threshold, must be included in a gauge-invariant
manner. This can be achieved by shifting the slepton mass into the
complex plane, $m_{\tilde{l}}^2 \to m_{\tilde{f}}^2 
- i m_{\tilde{l}} \Gamma_{\!\tilde{l}}$.
Moreover, for the production of off-shell sleptons 
the full $2\to4$ matrix element, including the decay of the
sleptons,  
must be taken into account 
as well as the  MSSM background and interference contributions.
One of the most important radiative corrections near threshold is the Coulomb
rescattering correction due to photon exchange between the slowly moving
sleptons.  
Beamstrahlung and ISR also play an important role.
The production of smuons and
staus proceeds via s-channel gauge-boson exchange, so that the
sleptons are produced in a P-wave with a characteristic rise of the
excitation curve $\sigma \propto \beta^3$, where $\beta = \sqrt{1-4
m_{\tilde{l}}^2/s}$ is
the slepton velocity. Due to the exchange of Majorana neutralinos in the
t-channel, selectrons can also be produced in S-wave ($\sigma \propto \beta$),
namely for $\seR^\pm \seL^\mp$ pairs in $e^+e^-$ annihilation and
$\seR^-\seR^-, 
\seL^-\seL^-$ pairs in $e^-e^-$ scattering.

Expectations for 
the R-selectron cross-sections at  both collider modes 
are shown in
Fig.\ref{fig:thrsel}
with the background from both the SM and
MSSM sources, reduced by appropriate cuts, included \cite{Freitas}.
Using five equidistant scan points, 
four free parameters, the 
mass, width, normalization and flat
background contribution, can be fitted 
in a model-independent way. 

\begin{figure}[h!]
\begin{minipage}[b]{6.5cm}
\epsfig{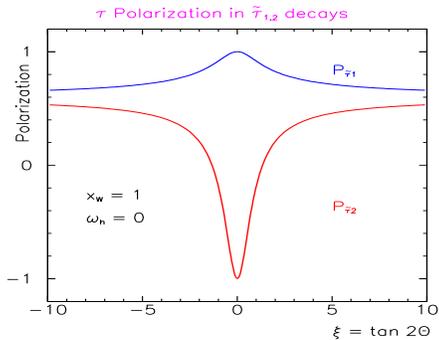} 
\end{minipage}%
\begin{minipage}[b]{6.5cm}
\caption{Polarization of the tau in the decay $\tilde{\tau}^-_{1,2} \to \tau^- 
\tilde{\chi}^0_1$ as a function of the stau mixing angle $\theta$
(which depends on  
$\tan\beta$) for bino-like $\tilde{\chi}^0_1$. From \cite{Boos:02}}
\label{fig:taupol}
\end{minipage}
\end{figure}

In contrast to the first two generation sfermions, large mixing are expected
between the left- and right-chiral components of the third generation sfermions
due to the large Yukawa coupling.
The mixing effects are thus sensitive to the Higgs parameters
$\mu$ and $\tan\beta$ as well as the trilinear couplings $A_{f}$
\cite{Nojiri}. 
For instance,  by examining the
polarization of the taus in the decay $\tilde{\tau}^-_{1,2} \to \tau^-_{\rm
L,R} \tilde{\chi}^0_1$, Fig.~\ref{fig:taupol}, a recent study within the MSSM
\cite{Boos:02} finds that $\tan\beta$ in the range of 30--40 can be
determined with an error of about 10\%.%

Moreover, if $A_{\tau}$ or $\mu$ turn out to be complex, 
the phase of the off-diagonal term 
$a_\tau m_\tau = ({\ A_\tau} -{\ \mu}^*\tan\beta)m_\tau
=|a_\tau m_\tau|{\rm e}^{i\Phi_\tau}$
modifies $\tilde{\tau}$ properties. 
Although the best would be to determine 
the complex parameters by measuring
suitable $CP$ violating observables, 
this is not straightforward, because the $\tilde \tau_i$
are spinless and their main decay modes are two--body decays. However, 
the $CP$ conserving observables also depend on the phases.

\begin{figure}[th!]
\begin{minipage}[b]{10.9cm}
\epsfig{figure=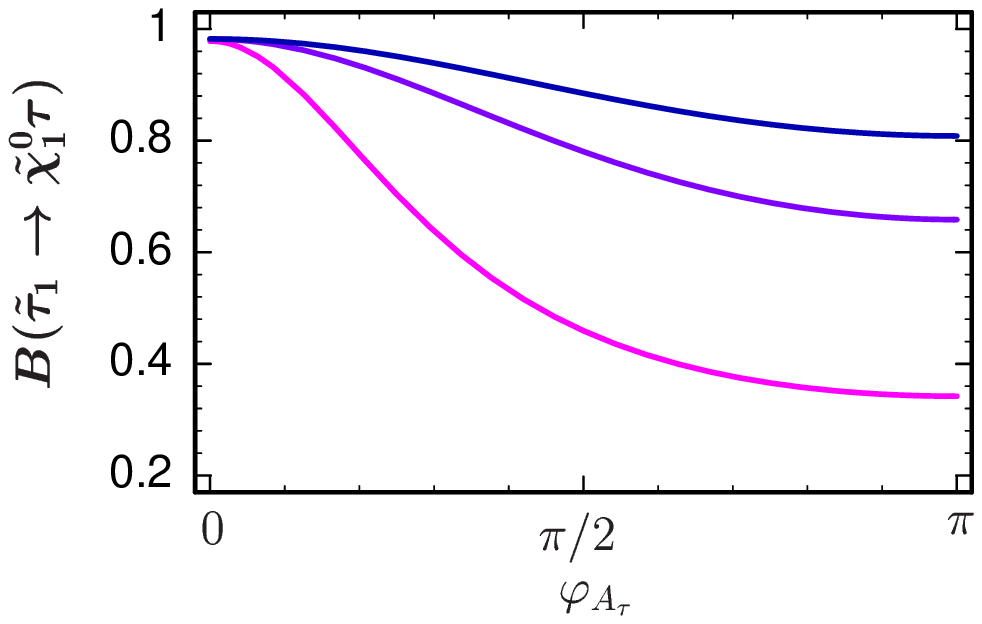, width=5.2cm,height=5cm}
\epsfig{figure=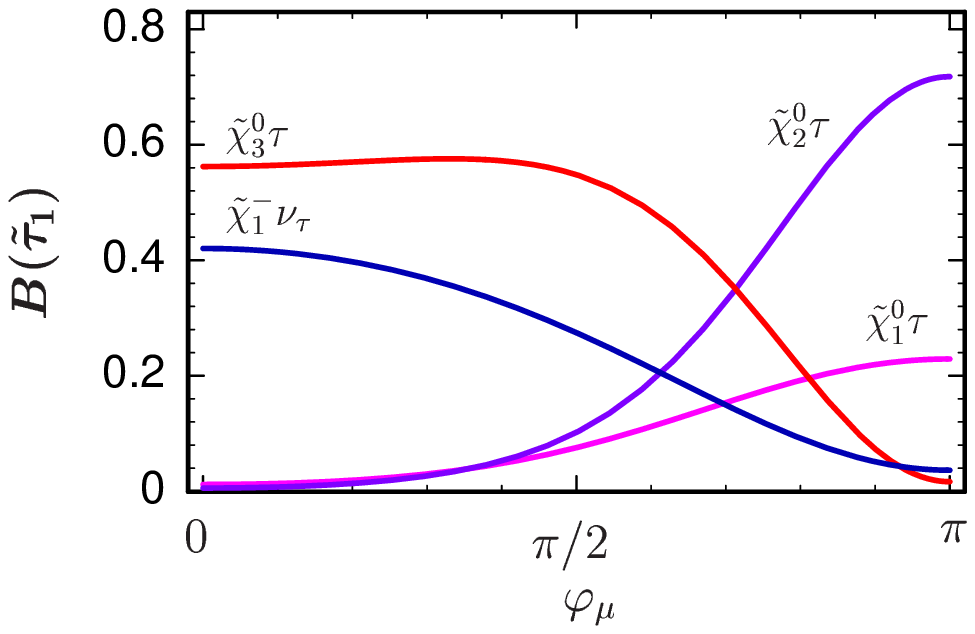, width=5.2cm,height=5cm}
\end{minipage}%
\begin{minipage}[b]{4.5cm}
\caption{Branching ratios of $\T_1$ and $\T_2$ as a function of:
$\varphi_{A_\tau}$ (left), $\varphi_\mu$ (right).   
From \cite{Bartlsl}. \label{taubr}}
\end{minipage}%
\end{figure}

For example, the various $\tilde\tau$ 
decay branching ratios depend in a
characteristic way on the complex
phases \cite{Bartlsl}. This is illustrated in Fig.~\ref{taubr}. 
The fit to the simulated experimental
 data with 2 ab$^{-1}$ at a collider like TESLA shows that  
 $\imag A_\tau$ and $\real
A_\tau$ can be determined with an error of order 10\%.

\begin{figure}[h!]
\begin{minipage}[b]{10.5cm}
 \epsfig{file=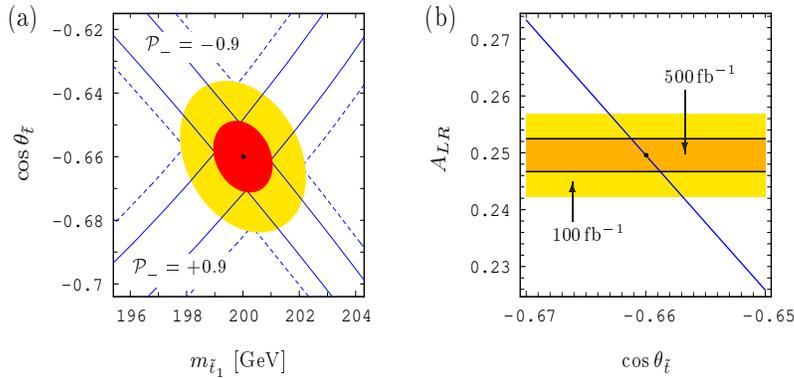,height=5cm,width=10.5cm}
\end{minipage}
\begin{minipage}[b]{5cm}
\caption{(a) Error bands and 68\% CL ellipse in 
\{$m_{\tilde{t}_1}$, $\cos\theta_{\tilde t}$\} plane 
from cross section measurements;  
${\int L}=100{\rm \ fb}^{-1}$ (dashes) and 
$500 {\rm \ fb}^{-1}$ (solid). 
(b) The determination of $\cos\theta_{\tilde t}$ from $A_{LR}$.
From \cite{Bartlstop}. \label{fig:stop}}
\end{minipage}
\end{figure}

Similarly, 
for the $\tilde{t}$ and  $\tilde{b}$ sectors, the  $L-R$ mixing 
can be important. By  
measuring the production cross sections with polarized beams
the squark masses and mixing angles can be  determined quite
precisely \cite{Bartlstop}, see Fig.\ref{fig:stop}.
For more details on the sfermion sector, we refer to \cite{consfer}


\section{Extrapolating to high-energy scale}

 Why we need high precision measurements?
The Standard Model physics is characterized by energy scales of order 100
GeV. However we expect the origin of supersymmetry breaking at the
high scale, near the Planck scale
$\Lambda_{\rm PL} \sim 10^{19}$~GeV or the grand unification [GUT] scale
$\Lambda_{GUT} \sim 10^{16}$~GeV.
Information on physics near the Planck scale may become available
from the well-controlled extrapolation of fundamental parameters 
measured with high precision at laboratory energies.
Although such extrapolations exploiting renormalization group
techniques 
extend over 13 to 16
orders of magnitude, they can
be carried out in a stable way in supersymmetric theories
\cite{witten:81}. 
Such a  procedure has very
successfully been pursued for the three electroweak and strong gauge
couplings providing the solid base of the grand
unification hypothesis. 

\begin{figure}[!t]
a)~~~ $M_j^{2}$ [GeV$^{2}$] \hspace{3.5cm}b)~~~ $M_j^2$ [GeV$^2$]\\
\put(150,-15){$Q$ [GeV]}
\put(350,-15){$Q$ [GeV]}
\epsfig{figure=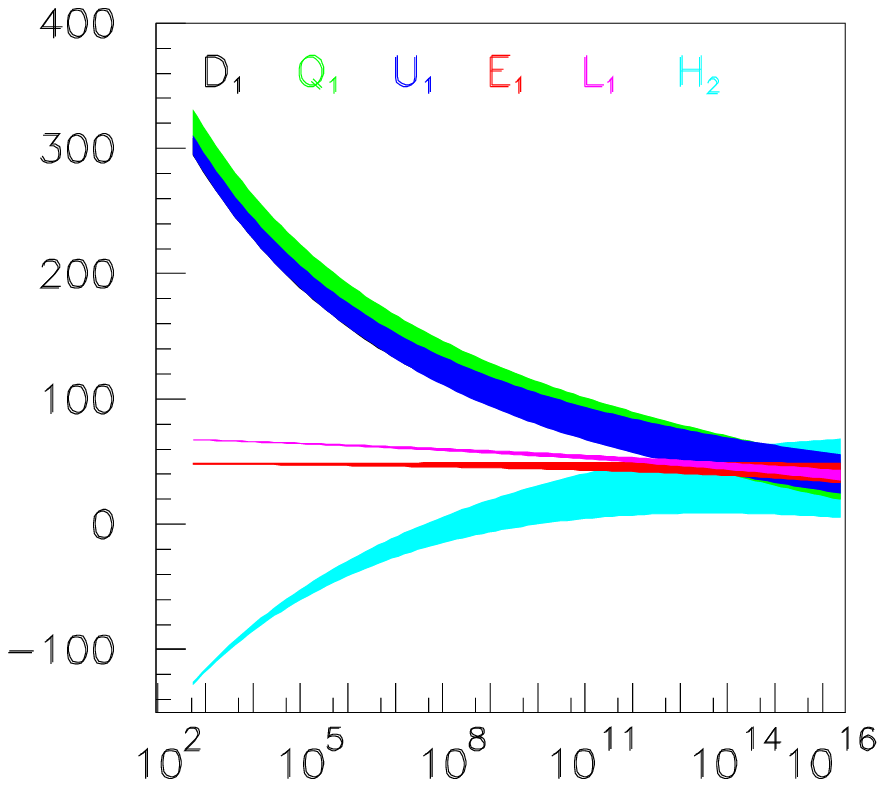,width=7cm,viewport=10 290 270 520}%
\epsfig{figure=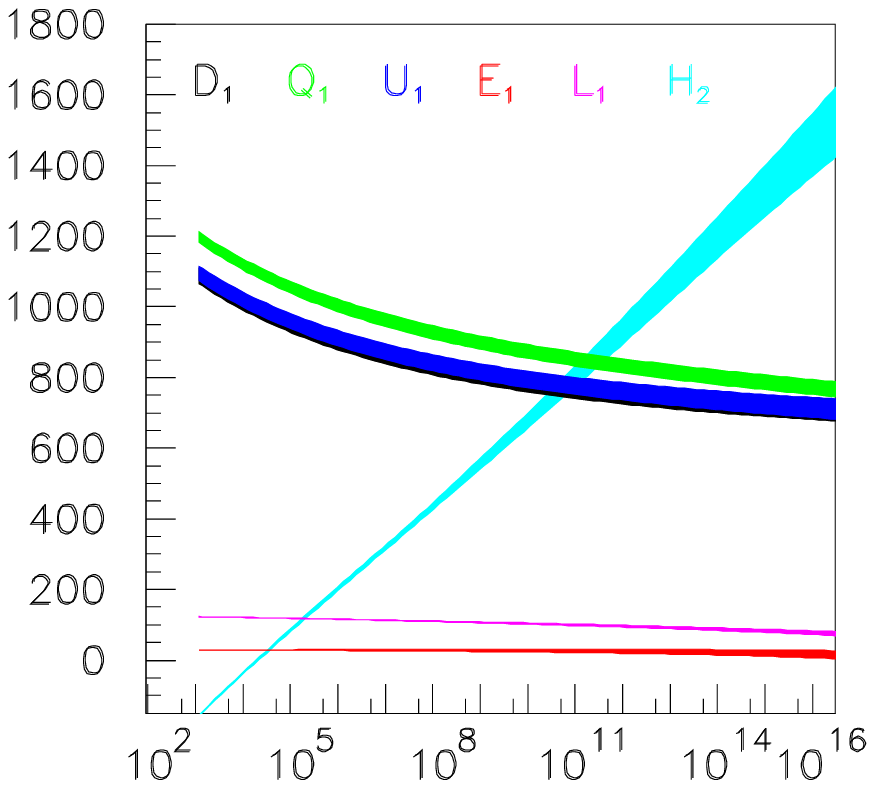,width=7cm,viewport=10 290 270 520}%
\caption{ Evolution, from low to high scales, of 
first-generation sfermion mass parameters
squared and the Higgs mass parameter $M^2_{H_2}$ for (a)   the mSUGRA
point SPS\#1a, (b)  the GMBS point SPS\#8. 
The widths of the bands indicate the 1$\sigma$ CL. From \cite{Blair}. 
\label{fig:extra} }
\end{figure}

This method can be expanded  to a large ensemble of 
the soft SUSY breaking parameters: 
gaugino and scalar
masses, as well as trilinear couplings. Recently  this procedure has
been applied \cite{Blair} to the  minimal supergravity
(with a naturally high degree of
regularity near the grand unification scale) and confronted with the gauge
mediated  supersymmetry breaking GMSB, see Fig.\ref{fig:extra}.

The basic structure in this approach is assumed to be essentially 
of desert type, although the existence of intermediate scales is not precluded.
An interesting example, the left-right extension of mSUGRA
incorporating the seesaw mechanism for the masses of right-handed
neutrinos, as well as a string-inspired effective field theory
example, 
can be found in \cite{Blair}. 

This bottom-up approach, formulated by means of
the renormalization group, makes use of the low-energy measurements to
the maximum extent possible. Therefore high-quality experimental data
are necessary in this context, that should 
become available by future lepton colliders, 
to reveal the
fundamental theory at the high scale.

\section{Conclusions}

Data rules! We need them badly. The LHC will provide plenty of data,
however, their theoretical interpretation will be possible in specific
models. In this context the $e^+e^-$ linear collider is very much
welcome. Overlap of the LC running with the LHC would greatly help to
perform critical tests: quantum numbers, masses, couplings etc.  We
have demonstrated that from the future high-precision data taken at
$e^+e^-$ linear colliders, TESLA in particular, and combined with
results from LHC, and CLIC later, the low-energy parameters of the
supersymmetric model can be determined.  Then the bottom-up approach,
by evolving the parameters from the low-energy scale to the high scale
by means of renormalization group techniques, can be exploited to
reconstruct the fundamental supersymmetry parameters at the high
scale, GUT or Planck, providing a
picture in a region where gravity is linked to particle physics,
and superstring theory becomes effective directly. 

\

\end{document}